# STM Studies of the Electronic Structure of Vortex Cores in $Bi_2Sr_2CaCu_2O_{8+d}$

S. H. Pan[1*], E.W. Hudson[1†], A.K. Gupta[2], K-W Ng[2], H. Eisaki[3‡], S. Uchida[3], and J.C. Davis[1]

[1] Department of Physics, University of California, Berkeley, CA, 94720-7300
[2] Department of Physics and Astronomy, University of Kentucky, Lexington, KY, 40506-0055.
[3] Department of Superconductivity, Tokyo University, 7-3-1 Hongo, Bunkyo-ku, Tokyo 113, Japan.
* Present Affiliation: Dept. of Physics, Boston University, Boston, MA 02215.
† Present Affiliation: Electron Physics Group, National Institute of Standards and Technology, Gaithersburg, MD 20899.
‡ Present Affiliation: Department of Applied Physics, Stanford University, Stanford, CA 94205-4060.

We report on low temperature scanning tunneling microscopy (STM) studies of the electronic structure of vortex cores in $Bi_2Sr_2CaCu_2O_{8+\delta}$ (BSCCO). At the vortex core center, an enhanced density-of-states (DOS) is observed at energies near $\Omega = \pm 7$ meV. Spectroscopic imaging at these energies reveals an exponential decay of these 'core states' with a decay length of $22 \pm 3$ Å. The four-fold symmetry sometimes predicted for d-wave vortices is not seen in spectroscopic vortex images. A locally nodeless order parameter induced by the magnetic field may be consistent with these measurements.

PACS# 74.72.Hs , 74.50.+r , 75.30.Hx , 61.16.Ch

In conventional type II superconductors, the order parameter (OP) is suppressed in the cores of quantized magnetic vortices [1] and recovers over a distance of about one coherence length $\xi$. Bound quasi-particle states can exist inside these cores [2] with lowest energy given approximately by $E \sim \Delta^2/2E_F$, where $E_F$ is the Fermi energy and $\Delta$ is the superconducting energy gap. Such 'core' states were first imaged by Hess *et al.* using low temperature scanning tunneling microscopy [3]. In the high temperature superconductors (HTSC) the zero magnetic field OP is believed to have mainly a $d_{x^2-y^2}$ symmetry [4-7] and is therefore not uniform in k-space but instead has four nodes [8]. The electronic structure of vortex cores in a superconductor with such an unconventional OP has attracted much attention, both theoretical [9-27] and experimental [28-31], since it serves as a coherence-length scale (~10 Å) probe of the condensate structure.

Evidence for low energy vortex core states in YBCO was first provided by infrared absorption experiments [28]. Pioneering STM experiments on YBCO by Maggio-Aprile *et al.* directly identified vortex core states at energies of ± 5.5 meV by tunneling spectroscopy [29]. Similar states were not however observed by STM in BSCCO vortex cores [30-31].

Here we report new STM experiments which demonstrate the existence of low energy states associated with the vortex cores in BSCCO. Two different types of as-grown BSCCO single crystals were used. The first set, grown by the directional solidification technique, has a transition temperature $T_c = 87$ K, a transition width of 5 K and contains scattering centers at a density of 0.3%, while the second set, grown by the floating zone technique, has a very dilute (~0.3%) doping concentration of Zn atoms acting as impurity scatterers, a $T_c$ of 84 K and a transition width of 4 K. All samples are cleaved in cryogenic ultra-high vacuum at 4.2 K and immediately inserted into the STM head [32].

In addition to topographic imaging, STM can be used to map the electronic DOS on the surface at energy E = eV by measuring the differential tunneling conductance $G$ at sample bias V as a function of position. When studied in zero magnetic field, images of only the impurity scattering resonances [33-35] appear in a zero-bias conductance map. Figure 1a is such an image and shows the impurity resonances as dark features with diameter ~ 30 Å. However, after application of a magnetic field perpendicular to the surface, additional features associated with the magnetic vortices appear in the zero-bias DOS map. An example of the tunneling spectrum measured at the center of such a feature is shown in Fig. 2, along with spectra from both a weak [33] and strong [35] impurity resonance, and from a 'regular' superconducting region [36].

Comparisons between the spectra in Fig. 2 reveal two remarkable phenomena. First, the spectrum at the vortex core center shows complete suppression of *both* coherence peaks at the superconducting gap edge, indicative of a strongly suppressed OP in this region. This suppression is highly local, with recovery of the coherence peaks occurring within only ~ 10 Å from the center. Furthermore it is not, as in conventional superconductors [3], accompanied by a complete transfer of spectral weight to a zero-bias conductance peak (ZBCP). Second, the spectrum measured at the vortex core center shows additional states with energies near $\Omega = \pm 7$ meV, as indicated by the arrows in Fig. 2. We refer to these states as "core states," without implication that they are necessarily of the Caroli-de Gennes type [2]. They are easily distinguishable as a spectral feature only within a few angstroms of the vortex center. Because of their small DOS magnitude (about 15% of the normal state density of states), the appearance of these states can be perturbed by the presence of even a weak impurity resonance.

The low energy spectra at the impurities and vortices are sufficiently distinct from each other that the scattering resonances and the vortex cores can be independently imaged. This can be done by taking simultaneous DOS maps at zero-bias and at the core state energy (±7 mV) respectively. For example, Fig. 1b is measured at V= 7 mV in a magnetic field of 7.25 Tesla on the identical area as Fig. 1a and contains regions of enhanced DOS with apparent radius near 60 Å, associated with vortices.

In a 215 nm square field of view, typically 170 ± 10 vortices (identified by 7 mV DOS mapping) appear when B = 7 T. Imaging such large numbers of vortices not only allows a statistically robust identification of the phase of the vortex solid, but also allows Voronoi diagram analysis [37] for the accurate determination of the flux $\phi$ associated with each vortex. For all field strengths and all crystals studied, the resulting value is $\phi = 2.1(1) \times 10^{-15}$ T m$^2$, consistent with a single magnetic flux quantum. Thus, DOS imaging *at the core state energy* efficiently identifies the vortices, and, conversely, this implies that the ±7 meV states are associated with *all* vortex cores in BSCCO.

The ability to independently map vortex and impurity locations allows one to study interactions between them. For example, by comparing Fig. 1a and 1b, one can see that in regions of high (low) impurity density the density of vortices is also high (low). This can be seen more clearly in the Zn doped crystals which have stronger scattering impurities. In Fig. 3 the positions of impurity states at Zn atoms (~ 30 Å diameter dark dots) are shown and the positions of vortex cores (identified by simultaneous +7 mV spectroscopic mapping) are overlaid as 60 Å diameter open circles. Fifty percent of the circles representing the vortices contain at least one impurity resonance. Since the probability of such a coincidence between uncorrelated distributions is at least four times smaller, apparently the Zn impurity atoms provide attractive potentials for pinning of the vortices and their random distribution is a source for the disorder in the vortex solid.

The simultaneous mapping technique also allows the study of vortex phenomena uninfluenced by impurities. To do so, we focus on vortices whose centers are more than 150 Å from the nearest impurity site. To further reduce the influence of the scattering resonances on the measurements of vortex core properties we study the +7 meV core states since this energy is farther form the typical scattering resonance energy of –1.5 meV.

For these isolated vortices, the radial dependence of the core state DOS, $G_{CS}(r)$, is found as follows. First, the measured differential conductance $G_M(r, f)$ is averaged over the azimuthal angle $f$ to give $G_M(r)$. The resulting curve is normalized to 1 at r = 0 to account for differences in measurement conditions between experimental runs (e.g. tip-sample separation). Then, an exponential decay [38] plus offset of the form $G_\infty + A\exp(-r/r_0)$ is fit to each $G_M(r)$ between $r = 10$ Å and $r = 75$ Å. The offset, $G_\infty$, is the expected non-zero DOS of a d-wave superconductor at finite energy (i.e. 7 meV). It is subtracted as a background from the data to yield the radial dependence of the additional core state DOS $G_{CS}(r) = G_M(r) - G_\infty$. Remarkably, for all vortices studied, the exponential fit parameters for $G_{CS}(r)$ are indistinguishable within the error bars, showing a decay length $r_0 = 22 \pm 3$ Å. We show, as open circles in Fig. 4, the averaged value of $G_{CS}(r)$ for 6 isolated vortices (three as indicated by arrows in Fig. 1b and three others from Zn-doped samples) along with the resulting exponential fit which shows good agreement over the whole range when $r > 10$ Å.

Similar analyses were also carried out on large numbers of vortices which are apparently pinned at impurity sites. For radii smaller than about 20 Å the $G_{CS}(r)$ vary dramatically due to the DOS perturbations from the impurity resonances. However, when the azimuthally averaged data from these pinned vortices (shown as solid triangles in Fig. 4) are fit to an exponential plus an offset for distances between 20 Å and 75 Å from the center of the vortex, their $G_{CS}(r)$ again all exhibit an exponential decay length of $22 \pm 3$ Å. This indicates that, at large distances from the pinning site, core states of pinned vortices are indistinguishable from those of unpinned vortices.

The apparent shape of the core state DOS varies, probably due to inhomogeneities in the material. To study the angular dependence of the core states we measure variations in $G_{CS}(r, f) = G_M(r, f) - G_\infty$ for each unpinned vortex. It is found that the variation of $G_{CS}(r)$ with $f$ can be as high as 25% of the averaged value of the fit $\langle G_F(r, f)\rangle$, but is random and has no reproducible four-fold symmetry and no preferential register to the gap node directions. Since such four-fold symmetric, low energy DOS features are seen at impurity scattering resonances in these Zn-doped samples [34], their absence here is significant.

The theoretical analyses of HTSC vortex cores contain discussions of several different scenarios. In the first, a BCS-type OP with pure $d_{x^2-y^2}$ symmetry results in "extended" quasi-particle states, characterized by a power law decay of their DOS with r, a four-fold symmetric shape oriented with the nodal directions, and a ZBCP at the core center [l2, 16, 17, 21]. This picture does not appear to be consistent with our observations. In another BCS-type scenario, a secondary OP component of a different symmetry is generated locally by the magnetic field and the resultant pair potential (with symmetry $d_{x^2-y^2}$ + is or $d_{x^2-y^2} + id_{xy}$ for example) no longer has nodes [10, 11, 14, 15, 17, 18, 21, 23, 26]. This can result in bound quasi-particle core states which decay exponentially with r. Due to the disappearance of the nodes, their DOS should be much more azimuthally symmetric than the extended states of the first scenario.

HTSC vortex cores have also been studied using non-BCS models. Predictions of insulating vortex cores in SO(5) theory [22], and in spin-charge separated models [25] are not apparently consistent with the observations. Some t-J model calculations have predicted a ZBCP split by an induced s-wave OP [19,20]. Recently a t-J model with Coulomb interaction has predicted weak, low energy core states, and their exponential decay [26], also in agreement with our observations.

When our measurements in BSCCO are compared with existing experimental STM data for vortex cores in YBCO [29], the observed phenomena appear to be quite similar. In both cases, there are no zero bias conductance peaks (as was previously reported for BSCCO [29-31]), particle-hole symmetric low energy core states appear at ± 5.5 mV (YBCO) and ± 7 mV (BSCCO), the vortex diameter is about 60 Å, and four-fold symmetry is not observed. Thus, it seems that a common phenomenology for vortices in the two major HTSC materials has now been identified.

In conclusion, we report the first observation of finite energy core states in BSCCO vortices and the first demonstration in any HTSC system of spectroscopic imaging of vortices by using the core states themselves. We show how spectroscopic imaging at the different energies characteristic of impurity states and vortices allows discrimination between their intrinsic properties and those due to interactions. The properties of the core state DOS appear not to be consistent with extended core states at zero energy in a pure $d_{x^2-y^2}$ OP. They may however be consistent with quasi-particle states bound in vortex cores whose order parameter is locally nodeless due to the magnetic field.

We acknowledge and thank A. V. Balatsky, A. J. Berlinsky, A. de Lozanne, M. Franz, N. B. Kopnin, D.-H. Lee, K. Maki, K. Moler, J. Orenstein, R.E. Packard, J. Sauls, E. Thuneberg, G. E. Volovik and S.-C. Zhang, for helpful conversations and communications. This work was supported by the NSF under grants DMR-9458015 and DMR-9972071, the LDRD Program of the Lawrence Berkeley National Laboratory under the Department of Energy Contract No. DE-AC03-76SF00098, the Packard Foundation, Grant-in-Aid for Scientific Research on Priority Area (Japan), and by a COE Grant from the Ministry of Education, Japan.

Figure 1. (a) A DOS map of a 120 nm square field of view, measured at zero sample bias in zero magnetic field, shows the impurity scattering resonances.
(b) A DOS map, measured on the same 120 nm square field of view as Fig. 1a at sample bias V = 7 mV in B = 7.25 T shows about 50 regions of increased DOS associated with vortices. The apparent shape of the vortices varies from vortex to vortex, probably due to the influence of impurities and other inhomogeneities in the crystal. The vortex solid appears highly disordered. Comparison with Fig. 1a shows that vortices are often found at the sites of impurities, which appear to act as pin sites.

Figure 2. Differential tunneling conductance spectra taken at different locations. The top two spectra, taken at the center of a Zn impurity resonance (strong) and an impurity resonance of unknown source (weak) respectively, show a peak in the DOS just below the Fermi energy (~ -1.5 mV). The third spectrum, taken on a 'regular' (free of impurity resonances and magnetic vortices) part of the surface, shows a superconducting energy gap with Δ = 32 mV. The bottom spectrum, taken at the center of a vortex core, shows two local maxima at ±7 mV, as indicated by the two solid arrows. In addition, both coherence peaks at the gap edge are completely suppressed. All curves were obtained at 4.2 K using a lock-in technique. The junction resistance was set to 1 GΩ at $V_{sample}$ = –200 mV, and the 447.3 Hz lock-in modulation had an amplitude of 500 $\mu V_{rms}$. The curves are offset by 0.75 nS for clarity.

Figure 3. A DOS map of a 215 nm square portion of a Zn-doped sample. This map, taken at V = 0, shows the location of approximately 430 Zn impurity states (which appear as roughly 30 Å diameter dark dots). The overlaid dark circles of diameter 60 Å represent vortices (about 170 in the field of view) whose locations were identified in a simultaneously obtained 7 mV DOS map. Data was obtained in a magnetic field of B = 7 T at T = 4.2 K, using the same lock-in parameters as in Fig. 2.

Figure 4. Semi-log plot of the azimuthally averaged differential conductance $G_{CS}(r) = G_M(r) - G_\infty$ at V = + 7 mV versus distance *r* from the center of a vortex. Two sets of data are displayed. The open circles are extracted from high-resolution conductance maps of six different vortices which were well isolated from impurity resonance sites. The solid triangles are from ten different vortices found centered at impurity sites. In both cases the differential conductance has been averaged over azimuthal angles, normalized, and then averaged over all data sets. The solid line is an exponential fit and gives a decay length of 22 ± 3 Å.

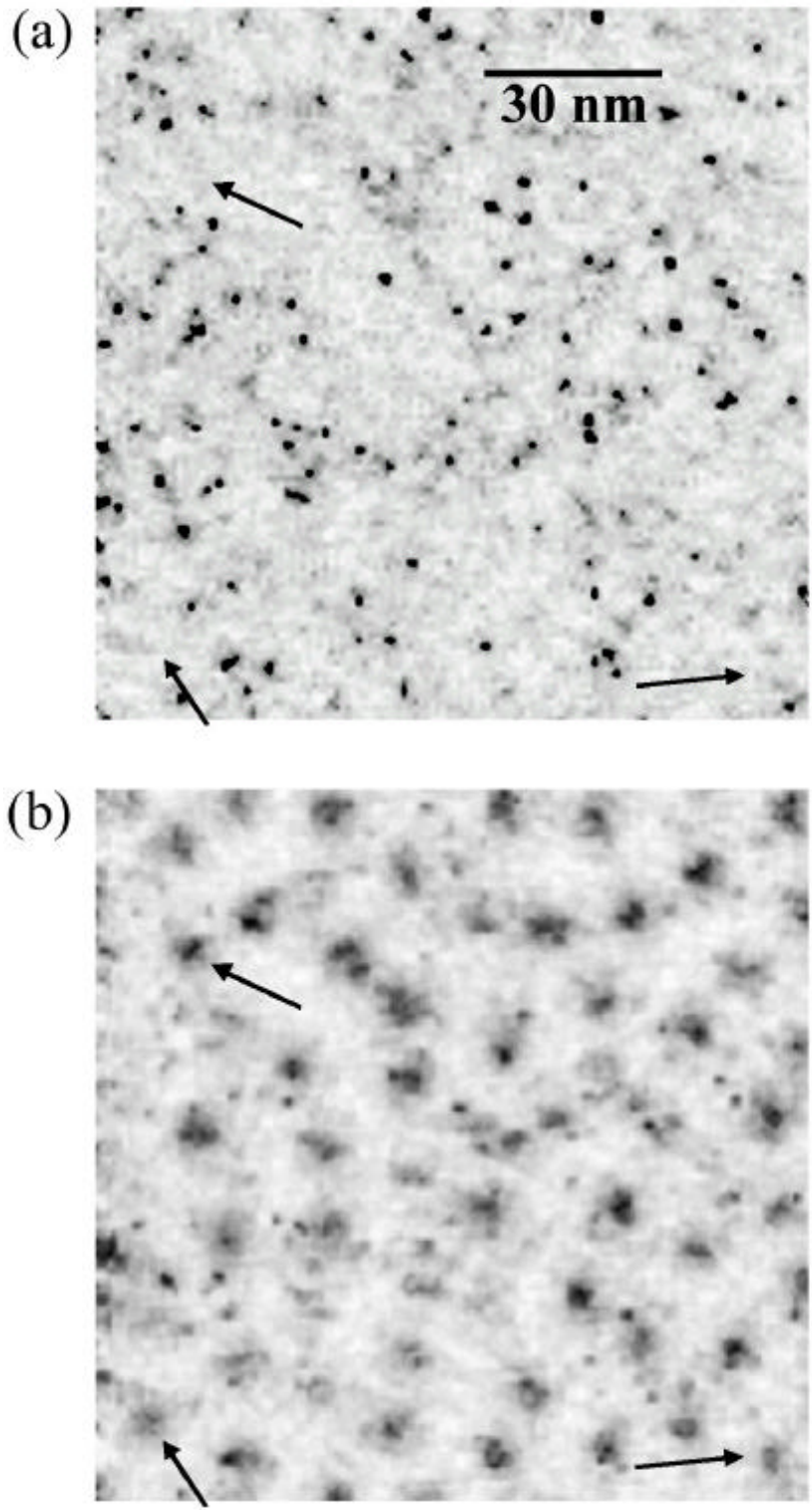

Figure 1
Physical Review Letters
S.H. Pan *et al*.

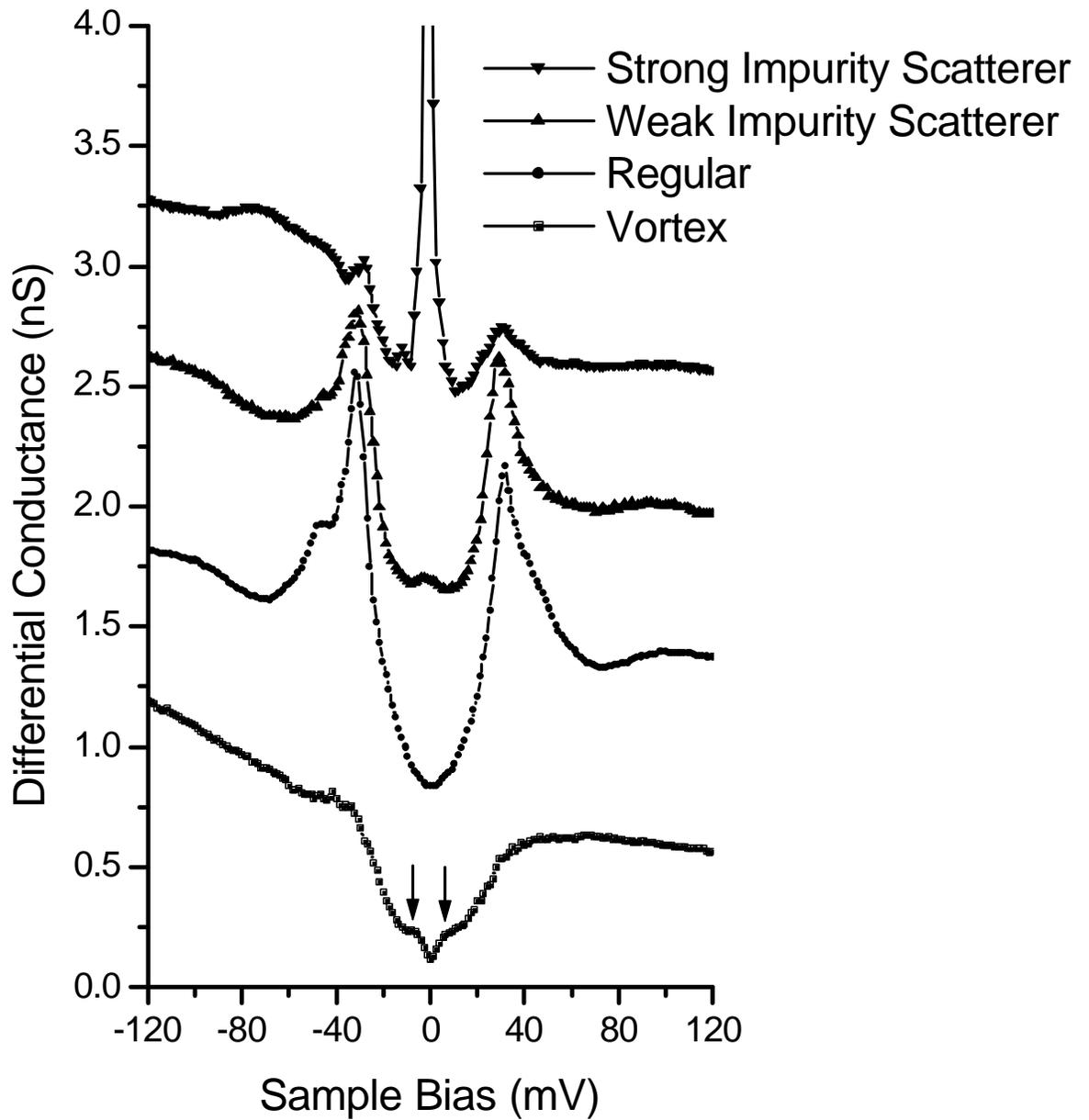

Figure 2
Physical Review Letters
S.H. Pan *et al.*

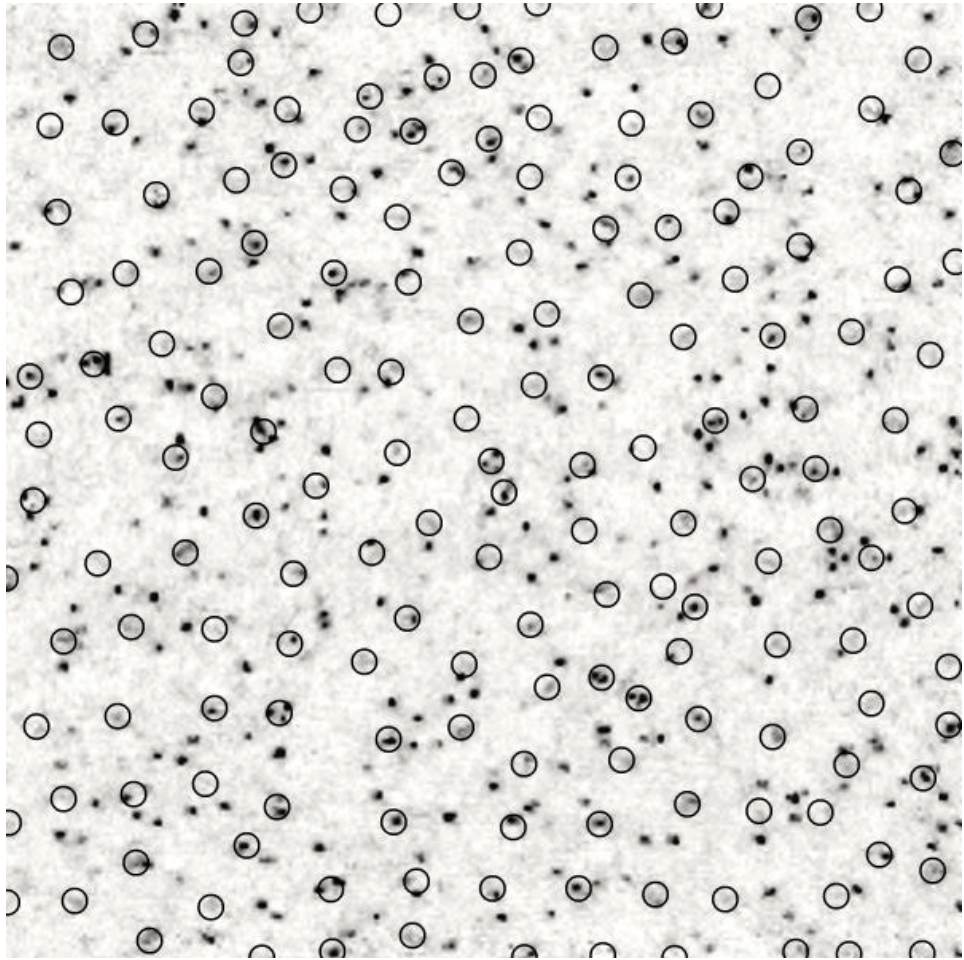

Figure 3
Physical Review Letters
S.H. Pan *et al.*

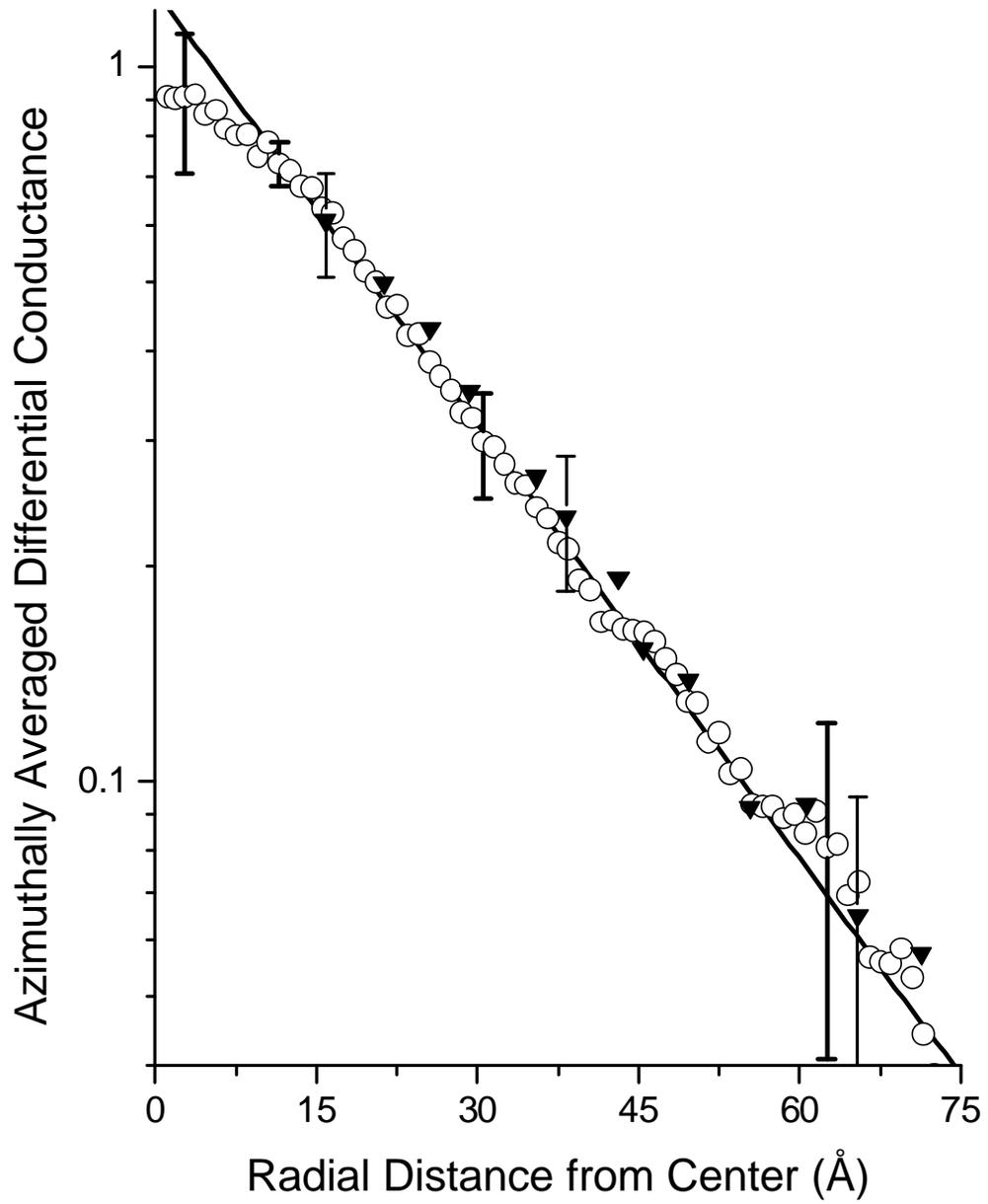

Figure 4
Physical Review Letters
S.H. Pan *et al.*